\documentstyle[12pt,epsf]{article}  
\begin{document}
\author{J. Zipprich, C. Fuchs, E. Lehmann, L. Sehn, S.W. Huang\\ and 
Amand Faessler \\ 
Insitut f\"ur Theoretische Physik der Universit\"at T\"ubingen,\\
Auf der Morgenstelle 14, D-72076 T\"ubingen, Germany}
\date{}
\title{Influence of the pion-nucleon interaction on\\ the collective
pion flow in heavy ion reactions}
\frenchspacing  
\maketitle  
\begin{abstract}
We investigate the influence of the real part of the in-medium pion optical 
potential on the pion dynamics in intermediate energy heavy ion 
reactions at 1 GeV/A. For different models, i.e. a phenomenological 
model and the $\Delta$--hole model, a pionic potential is extracted 
from the dispersion relation and used in Quantum Molecular Dynamics 
calculations. In addition with the inelastic scattering processes 
we thus take care of both, real and imaginary part of the 
pion optical potential. A strong influence of the real pionic 
potential on the pion in-plane flow is observed. In general such a potential 
has the tendency to reduce the anticorrelation of pion and nucleon 
flow in non-central collisions.
\vspace{5mm}
\end{abstract}  
\clearpage
\setcounter{page}{1}
\renewcommand{\footskip}{2.5cm}
The pions strongly dominate the production of mesons in 
intermediate heavy ion collisions. Due to their strong interaction 
with the nuclear medium it is, however, still an open question 
if pions can serve as a sensitive probe of hot and compressed 
nuclear matter. The analysis of pion spectra 
\cite{harris87,venema93,gillitzer96} 
is supposed to provide information about the creation and dynamics 
of resonaces in compressed and excited nuclear matter. Besides the 
dominant $\Delta$(1232) resonance high energetic pion spectra may 
also yield information about the role of higher resonances, e.g. the 
$N^*$(1440) \cite{Eh92}. In addition there have been also measurements 
of collective pion observables, i.e. the in-plane pion flow has 
been measured by the DIOGENE group for the system Ne+Pb at 
800 MeV/A \cite{diogene} and the squeeze out, i.e. the flow 
perpendicular to the reaction plane, by, e.g., the TAPS 
collaboration \cite{brill}.

Since the reaction dynamics are strongly 
influenced by the pionic channels of the NN inelastic collision 
processes the understanding of the pion dynamics is of particular 
interest. However, most theoretical transport calculations \cite{Li91,Ba95} 
included the interaction of the pions
with the surrounding nuclear medium only by collision processes, i.e.
parametrizing the imaginary part of the pion optical potential. 
Such calculations are able to reproduce qualitatively, e.g., the 
DIOGENE data but as a general feature seem to underpredict the total 
amount of flow \cite{Li91,hartnack88}. Hence, we present here 
studies which also include the real
part of the pion optical potential as a potential interaction for the pions
during their propagation through the nuclear medium and thus we take care 
of the full in-medium pion optical potential. 

The knowledge of the pion 
optical potential from elastic pion--nucleus scattering is, however, 
restricted to nuclear densities at and below saturation and relatively 
small energies \cite{SCM79}. In heavy ion 
reactions at intermediate energies 
from about 0.1 to 2 GeV/A baryon densities 
up to three times saturation density are reached and pion 
momenta of several hundred MeV can occure. In this range the 
real part of the pion potential is almost unknown. Thus, 
one has to extrapolate the pion dispersion 
relation to the ranges relevant for heavy ion collisions.
To obtain an estimation of such in-medium effects 
we apply two models, i.e. the 
$\Delta$--hole model \cite{We88,Fr81} and a phenomenological ansatz 
suggested by Gale and Kapusta \cite{Ga87}. The application 
of these models allows to investigate the influence of the 
possible boundary cases, i.e. a soft 
($\Delta$--hole) and a rather strong phenomenological 
potential, and to demonstrate the influence on 
pionic observable in heavy ion collisions, in 
particular on the pion in-plane transverse flow produced 
in non-central collisions.

The dynamics of the nucleus-nucleus collision are simulated
within the framework of Quantum Molecular Dynamics (QMD) and 
a soft momentum dependent Skyrme force is used for the 
nuclear mean field. A detailed description of the QMD approach can be 
found in Refs. \cite{Ai91,khoa}. We included the 
$\Delta(1232)$ and the $N^{*}(1440)$ baryonic resonances which 
as well as the pions originating from their decay are 
explicitely treated, i.e. in a non-pertubative way.
For the cross sections of the inelastic channels 
($NN\rightarrow N(\Delta,N^{*})$) and for the lifetimes 
of the resonances we use parametrizations 
determined from one-boson-exchange amplitudes in 
Born approximation \cite{Hu94}. For the decay of the $N^{*}$ and $\Delta$ 
following decay channels are taken into account: 
One pion decay ($\Delta,N^{*}\rightarrow N+\pi$), 
two pion decay ($N^{*}\rightarrow N+2\pi$) and the decay of the 
$N^{*}$-resonance in a $\Delta$-resonance and a pion 
($N^*\rightarrow\Delta+\pi$). These processes are calculated with energy 
dependent decay probabilities as given in Ref. \cite{Hu94}.
In all collision processes a medium dependence is included via the 
Pauli blocking of the final states according to the phase space occupancy. 
Furthermore, the momentum dependence of the nuclear mean field 
results in a (non-relativistic) effective mass of the baryons. 
Thus we are able to reasonably reproduce the total pion multiplicities, 
however, overestimate them by about 20\% which seems to be a common 
feature of present transport calculations \cite{Ba95}. This fact 
is supposed to be due to the description 
of the resonance rescattering channel 
($N(\Delta,N^{*})\rightarrow NN$) by a detailed balance argument 
which leads to an underestimation of this process \cite{Li91}. 
The multiplicities itself are found to be nearly unaffected by 
the real pionic potential. On the other handside, 
the pionic flow per particle does not react 
on slight changes in the multiplicities.

The pions are propagated between 
their collisions with the nucleons under 
consideration of the interaction with the surrounding nuclear matter. 
The interaction of the pions with surrounding nuclear 
matter results in a pion self energy $\Pi$ 
entering into the in-medium pion dispersion relation
\begin{eqnarray}
\omega({\bf q})^2= {\bf q}^2+m_{\pi}^2+\Pi(\omega,{\bf q},\rho)
\quad .
\label{1}
\end{eqnarray}
Here the self energy depends on the energy $\omega$ and the 
momentum ${\bf q}$ of the pion and on the nuclear matter density $\rho$. 
In the framework of the $\Delta$-hole model the self energy is 
finally given in the following form
\begin{eqnarray}
\Pi(\omega,{\bf q},\rho)=\frac{{\bf q}^2 
\chi(\omega,{\bf q},\rho)}{1-g'\chi(\omega,{\bf q},\rho)}
\label{2}
\end{eqnarray}       
with
\begin{eqnarray}
\chi(\omega,{\bf q},\rho)&=&-\frac{8}{9}
\left(\frac{f_{\Delta}}{m_{\pi}}\right)\rho
\frac{\omega_{\Delta}({\bf q})}{\omega_{\Delta}^2({\bf q})-\omega^2}
\nonumber\\
\omega_{\Delta}&=&\sqrt{M_{\Delta}^2+{\bf q}^2}-M_N
\quad .
\nonumber
\end{eqnarray}
The parameters entering into Eq. (\ref{2}), in particular 
the $\pi N\Delta$ coupling constant $f_{\Delta}$ 
and the correlation parameter $g'$ are taken in consistence 
with the OBE parameters of Ref. \cite{Hu94} and a consistent 
treatment of the real and imaginary part of the pion optical 
potential is achieved.

The self energy obtained from the $\Delta$-hole model, Eq. (\ref{2}), 
includes beside of excitation of $\Delta N^{-1}$ states 
also short range correlations of these states. In this 
approximation one neglects, however, terms of 
higher order \cite{Di87} which are necessary to prevent 
the system to undergo a phase transition to the so called pion 
condensation. Although this approach works 
reasonable at low baryon densities and low pion momenta \cite{Fr81} 
it yields the wrong boundary conditions when the pion passes through 
the surface of the nuclear matter into the vacuum. 
A possibility to avoid these unphysical features 
is to mix the two solutions of the dispersion relation, Eq. (1), i.e. the 
pion-like and the $\Delta$-hole-like branch in a way that 
the physical boundary conditions are fulfilled \cite{Eh92}. 
Proceeding this way pion condensation only takes place at unphysically 
high densities. However, such a construction leads to a strong 
softening of the in-medium effects and hence one can't be 
sure that the modified dispersion relation still represents the 
true pion-nucleon interaction.

To improve on this we also consider 
the phenemenological ansatz of Gale and Kapusta 
\cite{Ga87} which is motivated by the results of low-energy 
pion-nucleus scattering and the observation of pionic atoms. 
The pion dispersion relation reads
\begin{eqnarray}
\omega(q)&=& \sqrt{(|{\bf q}|-q_0)^2+m_0^2} - U 
\label{3}
\\
\mbox{with}\qquad U
&=&\sqrt{q_0^2+m_0^2}-m_{\pi}
\nonumber\\
m_0&=&m_{\pi}+6.5(1-x^{10})m_{\pi}
\nonumber\\
q_0^2&=&(1-x)^2m_{\pi}^2+2m_0m_{\pi}(1-x) 
\quad .
\end{eqnarray}
Here a phenomenological medium dependence is introduced 
via $x=e^{-a(\rho/\rho_0)}$ with the parameter $a=0.154$ 
which allows to extrapolate to higher densities thereby 
avoiding the appearence of pion condensation.

Fig. 1 shows the pion dispersion relation for two representative densities 
obtained by the two approaches. It becomes obvious that the 
phenemenological dispersion relation given by Eq. (\ref{3}) yields 
a strong attractive pion-nucleus optical potential (Pot.2) in comparison to 
the $\Delta$-hole model (Pot.1). In contrast to Pot.2 
the medium dependence of Pot.1 is relatively weak. 
Hence, the application of the two potentials allows 
to estimate the magnitude of such effects.

In Fig.2 the $\pi^-$ energy spectra for a central (b$\leq$2.8 fm) La+La 
reaction at 1.35 GeV/A are compared to the BEVELAC data of Ref. 
\cite{bevelac}. It is seen that for low energies the agreement with 
the experiment is generally improved by the inclusion of a 
pionic potential which, due to its attraction, lowers the kinetic energy of 
the pions. With increasing energy this behavior becomes more complex. 
The weak Pot.1 is still in a reasonable agreement with the data. 
In the case of Pot.2 the attraction is, however, so pronounced that the 
pions are forced to follow the trajectories of the nucleons resulting in 
an enforcement of the high energy components of the spectrum.  Since the 
high energy tails of, e.g., $p_t$-spectra measured by the TAPS collaboration 
\cite{venema93} are systematically underestimated by conventional 
calculations \cite{Ba95} the inclusion of a pionic potential may 
help to resolve on this problem.

A quantity of particular interest is the pion transverse flow. 
In previous studies an anticorrelation 
of the pion and the nucleon transverse flow 
has been predicted for non-central collisions \cite{Ba95}. 
New results of the FOPI Collaboration seem to confirm 
this observation \cite{Pi95}. 
Such an anticorrelation which in part has also been observed by 
the DIOGENE group is explained by a shadowing 
effect due to the absorption and rescattering of the pions by 
spectator nucleons. Due to the large asymmetry of the considered system 
the DIOGENE data are, however, strongly distorted by the 
participant-spectator geometry and no definit (anti)correlation 
has been observed over the entire rapidity range of the 
reactions. The magnitude of this shodowing 
effect grows with the number and the flow of the 
spectator nucleons, i.e. it is proportional to the impact parameter. 
This can be seen from Fig.3 where the directed transverse momentum 
$p^{dir}_x = \frac{1}{N} \sum_i sign(Y_{CM}^i) p_{x}^i$ 
is shown as a function of the impact parameter 
for the system Ca+Ca at 1 GeV/A. The negative directed flow 
in non-central collisions indicates the anticorrelation 
with the positive directed flow of the nucleons. It is further seen 
that the attractive potential interaction in general has the tendency to 
reduce this anticorrelation and a strong potential like Pot.2 
can even convert it into a correlation.

This behaviour can be explained as follows: 
the real part of the pion optical potential leads to a bending of the 
trajectories of the pions in the direction 
where the majority of the nucleons moves. The two effects, i.e. 
the potential and the shadowing effect, counterbalance each other and hence, 
if the pion-nucleus potential is strong enough one observes even a 
correlated pion flow.

Annother question of interest is the influence of the nuclear equation of state 
(EOS) on the pion flow. The previous results have been obtained 
with a momentum dependent Skyrme force \cite{khoa} 
corresponding to a soft EOS. Fig.4 compares to 
the pion in-plane flow obtained with a momentum independent soft Skyrme and a 
momentum dependent hard Skyrme force. It turns out that the pion in-plane flow 
does not react very sensitiv on the nuclear EOS compared to the strong 
influence of the pion optical potential and attempts to fix the 
EOS from there as suggested, e.g., in Ref. \cite{Ba95} will 
remain ambigous as long as these effects are not taken into account 
properly.

We conclude that the effect of the pion-nucleus optical potential 
extracted from the in-medium pion dispersion relation is of 
essential importance for the pion transverse flow. The dependence on the 
nuclear EOS is relatively small. This fact opens 
the possibilty to extract some information about the structure of 
the pion dispersion relation, in particular for its unknown ranges, 
from the pion flow in heavy ion collisions. However, in our opinion 
no final conclusions can be drawn from the present pion flow data. 
To obtain quantitative 
statements more refined measurements 
are necessary, in particular a detailed analysis of 
forthcoming results from the FOPI collaboration may 
help to clarify this situation.

\begin{center}
\begin{figure}\vspace*{3cm}
\leavevmode
\epsfxsize=13cm
\epsffile[30 85 430 410]{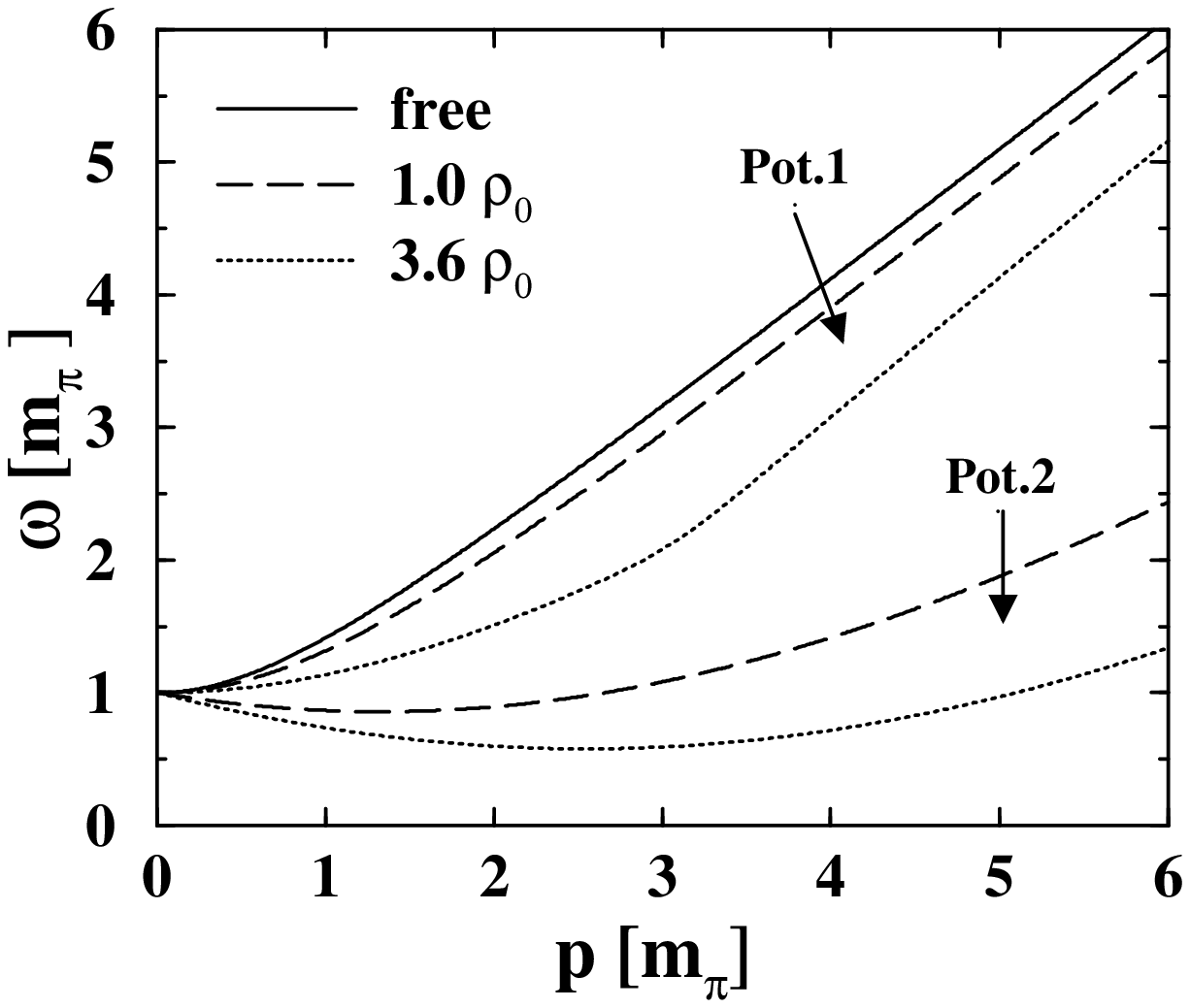}
\caption{
Comparison of the pion in-medium dispersion relations obtained 
within the $\Delta$-hole model (Pot.1) and with the phenomenological 
ansatz according to Ref. \protect\cite{Ga87} (Pot.2).
}
\end{figure}
\end{center}
\begin{center}
\begin{figure}\vspace*{3cm}
\leavevmode
\epsfxsize=13cm
\epsffile[30 85 430 410]{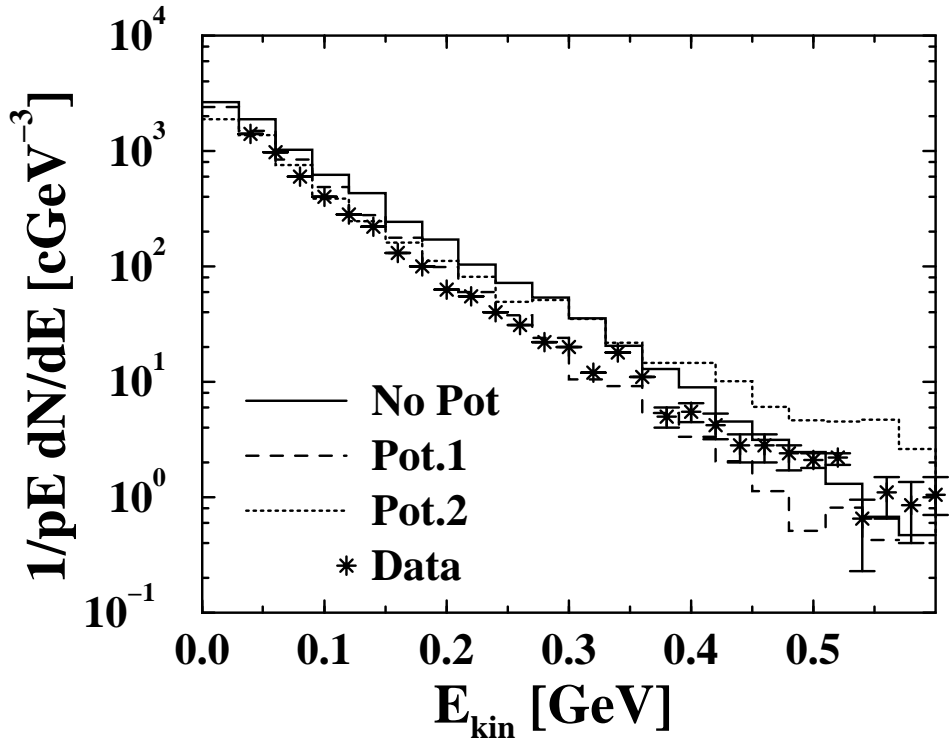}
\caption{
Comparison of $\pi^-$ energy spectra 
for a central ($b\leq 2.8$fm) La+La reaction at 1.35 GeV/A. The 
calculations have been performed with and without 
the real part of the pion-nucleus optical potential. The data 
are taken from Ref. \protect\cite{bevelac} and an angular cut 
of $60^{\circ} \leq \Theta_{CM} \leq 120^{\circ}$ has been 
applied.
}
\end{figure}
\end{center}
\begin{center}
\begin{figure}\vspace*{3cm}
\leavevmode
\epsfxsize=13cm
\epsffile[30 85 430 410]{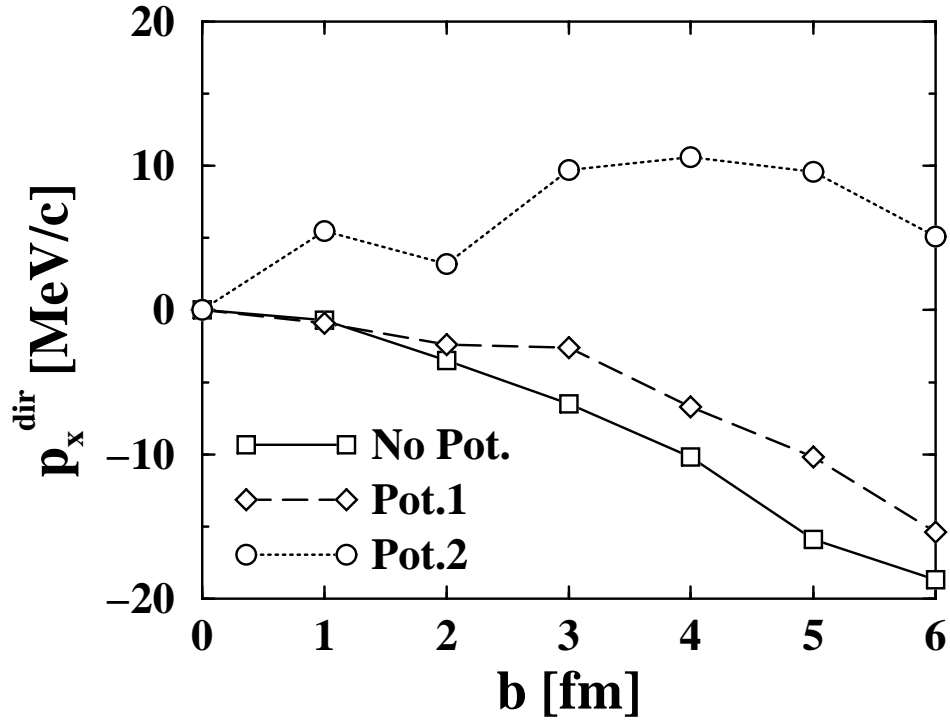}
\caption{
 Directed pion flow $p_x^{dir}$ as a function 
of the impact parameter with and without the real part of 
the pion-nucleus optical potential for the system Ca+Ca 
at 1 GeV/A.
}
\end{figure}
\end{center}
\begin{center}
\begin{figure}\vspace*{3cm}
\leavevmode
\epsfxsize=13cm
\epsffile[30 85 430 410]{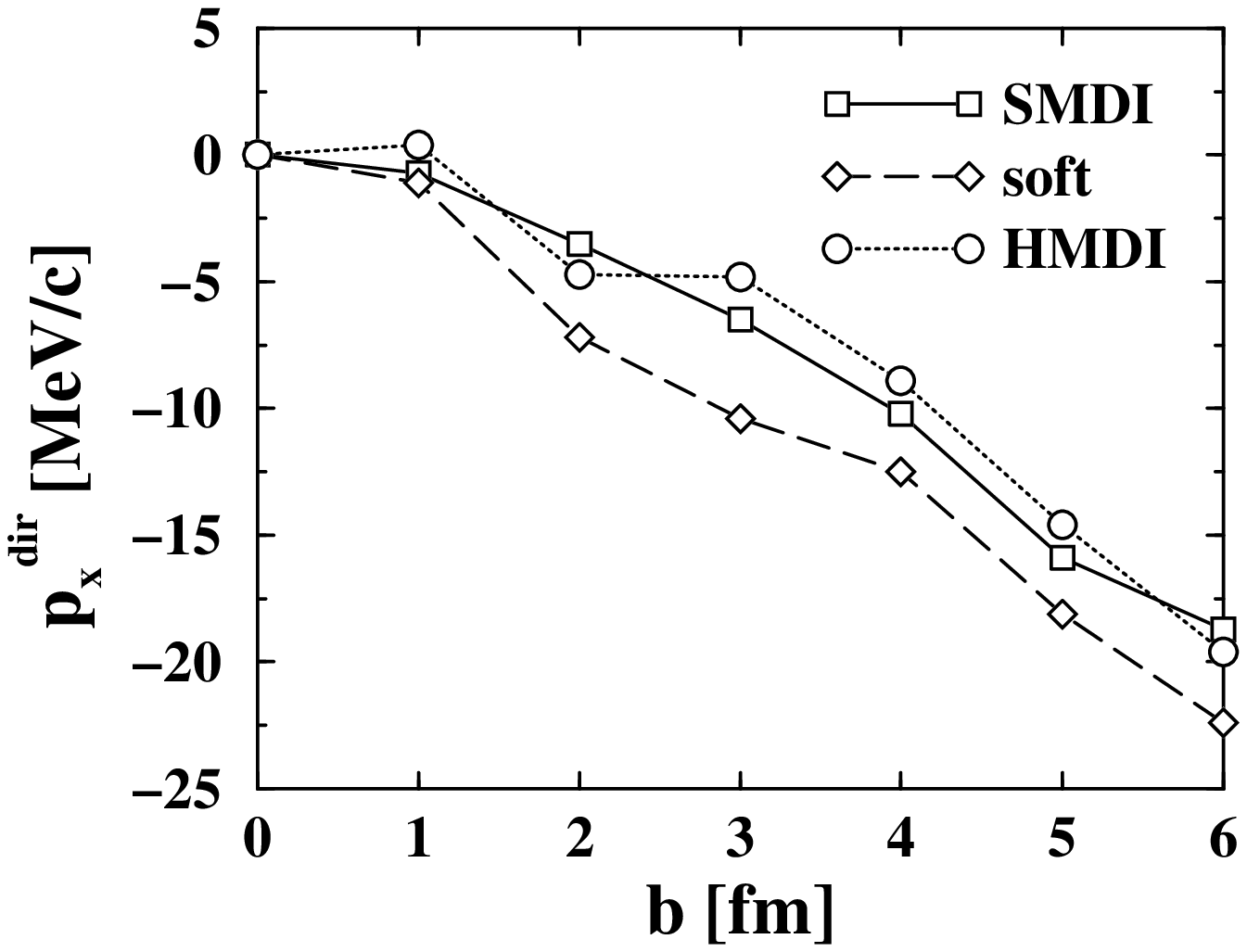}
\caption{
Dependence of the directed pion flow $p_x^{dir}$ 
as a function of the impact parameter on the nuclear 
equation of state. The calculations are performed for the 
same system as in Fig.3 using a soft/hard momentum dependent 
Skyrme force (SMDI/HMDI) and a soft Skyrme force 
without momentum dependence (dashed line). In these calculations 
the real part of the pion optical potential has not been taken 
into account.
}
\end{figure}
\end{center}

\newpage
\begin{thebibliography}{99}
\bibitem{harris87}
J.W. Harris et al., Phys. Lett. {\bf B153} (1987) 463.
 
\bibitem{venema93}
L. Venema and the TAPS Collaboration, 
Phys. Rev. Lett. {\bf 71} (1993) 835.
 
\bibitem{gillitzer96}
A. Gillitzer et al., Z. Phys. {\bf 354} (1996) 3.

\bibitem{Eh92}
W. Ehehalt, W. Cassing, A. Engel, 
U. Mosel and Gy. Wolf, Phys. Lett. {\bf B298} (1992) 31. 

\bibitem{diogene}
J. Gosset and the DIOGENE Collaboration, 
Phys. Rev. Lett. {\bf 62} (1989) 1251.

\bibitem{brill}
D. Brill et al., Phys. Rev. Lett. {\bf 71} (1993) 336.

\bibitem{Li91}
Bao-An Li, W. Bauer and G.F. Bertsch, 
Phys. Rev. {\bf C44} (1991) 2095.

\bibitem{Ba95}
S.A. Bass, C. Hartnack, 
H. St\"ocker and W. Greiner, Phys. Rev. {\bf C51} (1994) 3343.

\bibitem{hartnack88}
C. Hartnack, H. St\"ocker and W. Greiner, 
Proceedings on the 'International Workshop on Gross Properties 
of Nuclei and Nuclear Excitations XVI', 
Hirschegg, 1988, ed. by H. Feldmeier.

\bibitem{SCM79}
K. Stricker, H. McManus and J.A. Carr, 
Phys. Rev. {\bf C19} (1979) 929.

\bibitem{We88}
T. Ericson and W. Weise, Pions and Nuclei, 
Carendon Press Oxford 1988.

\bibitem{Fr81}
B. Friedmann, V.R. Pandharipande 
and Q.N. Usmani, Nucl. Phys. {\bf A372} (1981) 483.

\bibitem{Ga87}
C. Gale and J. Kapusta, Phys. Rev. {\bf C35} (1987) 2107.

\bibitem{Ai91}
J. Aichelin, Phys. Rep. {\bf 202} (1991) 233.

\bibitem{khoa} 
D.T. Khoa, N. Ohtsuka, M.A. Matin, A. Faessler, S.W. Huang,
E. Lehmann and R.K. Puri, Nucl. Phys. {\bf A548} (1992) 102.

\bibitem{Hu94}
S. Huber and J.Aichelin, Nucl. Phys. {\bf A573} (1994) 587.

\bibitem{Di87}
W.H. Dickhoff and H. M\"uther, Nucl. Phys. {\bf A473} (1987) 394.


\bibitem{bevelac}
C. Odyniec et al., LBL Report {\bf 24580} (1988) 215.

\bibitem{Pi95}
C.H. Pinkenburg, thesis, GSI-Darmstadt (1995).








\end{thebibliography}
\end{document}